\documentclass[a4paper,fleqn]{cas-dc}

\usepackage[authoryear]{natbib}
\usepackage{graphicx}
\usepackage{float}
\usepackage{subcaption}
\usepackage{url}
\usepackage{nccmath}
\usepackage{tabularx} 
\usepackage{booktabs}
\usepackage{setspace}

\def\tsc#1{\csdef{#1}{\textsc{\lowercase{#1}}\xspace}}
\tsc{WGM}
\tsc{QE}
\tsc{EP}
\tsc{PMS}
\tsc{BEC}
\tsc{DE}

\newif\ifreview 
\reviewfalse

\begin{document}
\let\WriteBookmarks\relax
\def\floatpagepagefraction{1}
\def\textpagefraction{.001}
\shorttitle{Energy Efficiency in Cloud-Based Big Data Processing for Earth Observation: Gap Analysis and Future Directions}
\shortauthors{A. Bhawiyuga et~al.}

\title [mode = title]{Energy Efficiency in Cloud-Based Big Data Processing for Earth Observation: Gap Analysis and Future Directions}

\author[1]{Adhitya Bhawiyuga}[orcid=0000-0003-1008-2524]
\ead{a.bhawiyuga@utwente.nl}
\credit{Conceptualization, Methodology, Formal analysis, Investigation, Data Curation, Writing - Original Draft, Writing - Review \& Editing, Visualization}
\author[1]{Serkan Girgin}[orcid=0000-0002-0156-185X]
\ead{s.girgin@utwente.nl}
\credit{Conceptualization, Methodology, Writing - Review \& Editing, Supervision}
\author[1]{Rolf A. de By}[orcid=0000-0002-8921-2960]
\ead{r.a.deby@utwente.nl}
\credit{Conceptualization, Writing - Review \& Editing, Supervision}
\author[1]{Raul Zurita-Milla}[orcid=0000-0002-1769-6310]
\ead{r.zurita-milla@utwente.nl}
\credit{Conceptualization, Writing - Review \& Editing, Supervision}

\address[1]{Department of Geo-Information Processing, Faculty of Geo-Information Science and Earth Observation, University of Twente \\ Langezijds Building, 
Hallenweg 8,
7522 NH Enschede, The Netherlands}

\begin{abstract}
    Earth observation (EO) data volumes are rapidly increasing. While cloud computing are now used for processing large EO datasets, the energy efficiency aspects of such a processing have received much less attention. This issue is notable given the increasing awareness of energy costs and carbon footprint in big data processing, particularly with increased attention on compute-intensive foundation models. In this paper we identify gaps in energy efficiency practices within cloud-based EO big data (EOBD) processing and propose several research directions for improvement. We first examine the current EOBD landscape, focus on the requirements that necessitate cloud-based processing and analyze existing cloud-based EOBD solutions. We then investigate energy efficiency strategies that have been successfully employed in well-studied big data domains. Through this analysis, we identify several critical gaps in existing EOBD processing platforms, which primarily focus on data accessibility and computational feasibility, instead of energy efficiency. These gaps include insufficient energy monitoring mechanisms, lack of energy awareness in data management, inadequate implementation of energy-aware resource allocation and lack of energy efficiency criteria on task scheduling. Based on these findings, we propose the development of energy-aware performance monitoring and benchmarking frameworks, the use of optimization techniques for infrastructure orchestration, and of energy-efficient task scheduling approaches for distributed cloud-based EOBD processing frameworks. These proposed approaches aim to foster more energy awareness in EOBD processing , potentially reducing power consumption and environmental impact while maintaining or minimally impacting processing performance.
\end{abstract}

\begin{keywords}
earth observation \sep big data \sep cloud computing \sep energy efficiency \sep gap analysis
\end{keywords}

\maketitle

\printcredits

\section{Introduction}
    Earth observation (EO) has long been crucial for monitoring and understanding our planet. Advances in aerospace technology like satellites and low-altitude aircraft have significantly increased the volume and heterogeneity of EO big data (EOBD), including spatial, temporal, and spectral information. This, in turn, has enabled more applications such as disaster impact assessment, resource mapping, and climate change studies~\citep{kalantar2020landslide, feng2023mapping, hu2018climatespark, guo2015earth}. Yet, the management and processing of large-scale data, particularly at national or global levels analysis, continues to present challenges and demands efficient strategies for data storage and computing resources~\citep{echterhoff2021earth,arab2022}.

    Cloud computing has become essential for processing the increasing volume and heterogeneity of EOBD and offers dynamic scaling and resource allocation to support distributed processing capability. Public cloud services like Google Earth Engine, Copernicus Data Space Ecosystem, and Pangeo Cloud offer extensive EO imagery repositories and computational platforms for large-scale analysis~\citep{zhao2021progress, jutz2020copernicus, lukacz2022data, karra2021global}. These platforms streamline complex analyses, making them more accessible to researchers and developers in the EO domain. Additionally, open-source cloud platforms like Pangeo allow organizations to deploy similar services within their own data centers, thereby increasing flexibility and control over workflows~\citep{abernathey2021cloud}. As a consequence, these cloud-based platforms are increasingly being adopted by the EO community for spatiotemporal data processing and visualization.

    This increased reliance on cloud computing for EOBD processing raises concerns about its energy consumption and carbon footprint. As of 2024, data centers already consume 1-4\% of total electricity in major economies like the United States, China, and the European Union~\citep{kamiya2025data}. In some nations, the impact is even more pronounced. In Ireland for example, data centers are expected to account for over 20\% of the country's total energy consumption. This demand is accelerating rapidly as the global data center energy use has surged by approximately 80\% since 2018~\citep{kamiya2025data}. To address the impact of these demands, various policy measures have been introduced. For example, the city of Amsterdam temporarily banned new data center construction from 2019 to 2020 and later limited data center power capacity to 670~MVA until 2030~\citep{saltzman2023unraveling}. Similarly, other regions have implemented energy efficiency standards and incentives for green data centers. These measures underscore the urgency of the problem at the infrastructure level, yet the efficiency of big data processing is equally dependent on the energy awareness of its software stack.

    In this regard, while energy consumption issues have gained attention in some big data domains, especially in artificial intelligence (AI)~\citep{schwartz2020green}, the EO community has been slower to address energy-related issues. Researchers in AI domain now often report the energy costs and carbon emissions of training and deploying large models~\citep{samsi2023words, patel2024characterizing}. In contrast, such considerations are overlooked in the EOBD domain. This oversight is particularly notable given the scale of EOBD operations, which involve petabytes of data and continuous satellite feeds~\citep{nativi2015big}. The absence of energy consumption reporting in EOBD workflows makes it difficult to evaluate and compare their environmental impacts. Additionally, while cloud providers claim their data centers are efficient, they offer little transparency about the specific energy costs tied to storing data and running computation workloads~\citep{bharany2022systematic}. This gap not only limits efforts to enhance energy efficiency but may also create challenges for aligning EOBD applications with their intended environmental goals~\citep{lisboa2024earth}. As the EO community increasingly relies on powerful cloud computing, adopting practices from other big data fields and developing EOBD-specific frameworks for measuring and reducing energy and carbon costs becomes crucial~\citep{SONG201980}.

    This paper explores the energy efficiency approaches in EOBD processing, identifies the key challenges and proposes strategic enhancements. The main contributions of this paper are as follows:
    \begin{enumerate}
    \item We identify and analyze the general characteristics of EOBD, including its data access patterns and processing task dependencies alongside with its impact to the energy consumption (Section \ref{section:eobd_characteristics}).

    \item We provide a comprehensive overview of the current state of big data services in EOBD domain, with a particular focus on the prevalent use of cloud computing platforms for EOBD storage and processing (Section \ref{section:landscape}).

    \item We assess common approaches to improve energy efficiency in various big data environments (Section \ref{section:approach}).

    \item We delineate the gaps in energy efficiency practices associated with EOBD processing on cloud platforms (Section \ref{section:challenges}).

    \item We propose areas for future EODB energy-related studies, including the creation of benchmarking frameworks and energy measuring tools specifically designed for EOBD workflows, and the development of optimization strategies for infrastructure orchestration and task scheduling in cloud-based processing. (Section \ref{section:improvement}).
    \end{enumerate}
    By identifying critical research gaps and outlining future directions, this study aims to lay the groundwork for more energy-efficient EOBD processing.

\section{Characteristics and energy implications of EOBD processing}\label{section:eobd_characteristics}
    The processing of EOBD is shaped by a combination of data characteristics, access patterns, and workflow structures, each presenting challenges for energy efficiency.
    
    \subsection{The scale of EOBD}
    As of 2024, over 900~active EO satellites equipped with various sensors are in orbit, including optical imaging, multispectral/hyperspectral imaging, radar, infrared, and other purposes~\citep{wilkinson2024environmental}. Collectively, these satellites have produced large amounts of EO data, with combined archieves on platforms like ESA Copernicus and NASA EOSDIS exceeding 800 PB in 2023 and growing by approximately 100 PB annually~\citep{wilkinson2024environmental}. From a velocity perspective, the exponential growth stems from multiple drivers including a growing number of orbiting satellites, sensor enhancement with higher resolution, and shorter revisit time~\citep{yao2023comparison}. Furthermore, the data exhibits significant variety. Each mission's unique sensors and orbital cycles result in data with various spatial, temporal, and spectral resolutions~\citep{zhao2023srsf, qian2021hyperspectral, phiri2020sentinel}.

    This scale of EOBD directly translates into the energy consumption in three ways. From data storage perspective, maintaining scalable and reliable petabyte-scale archives requires data centers where storage arrays and the essential cooling systems consume power continuously. The ongoing data transmission between satellites, ground stations, archive facilities, processing clusters, and end users, consumes substantial energy, particularly within networking infrastructure. Finally, the diverse data formats and resolutions demand pre-processing operations that utilize significant CPU or GPU cycles to generate analysis-ready data for end users.

    \subsection{Data access patterns}

\begin{figure}
    \centering
    \begin{subfigure}{0.3\textwidth}
        \centering
        \includegraphics[width=\textwidth]{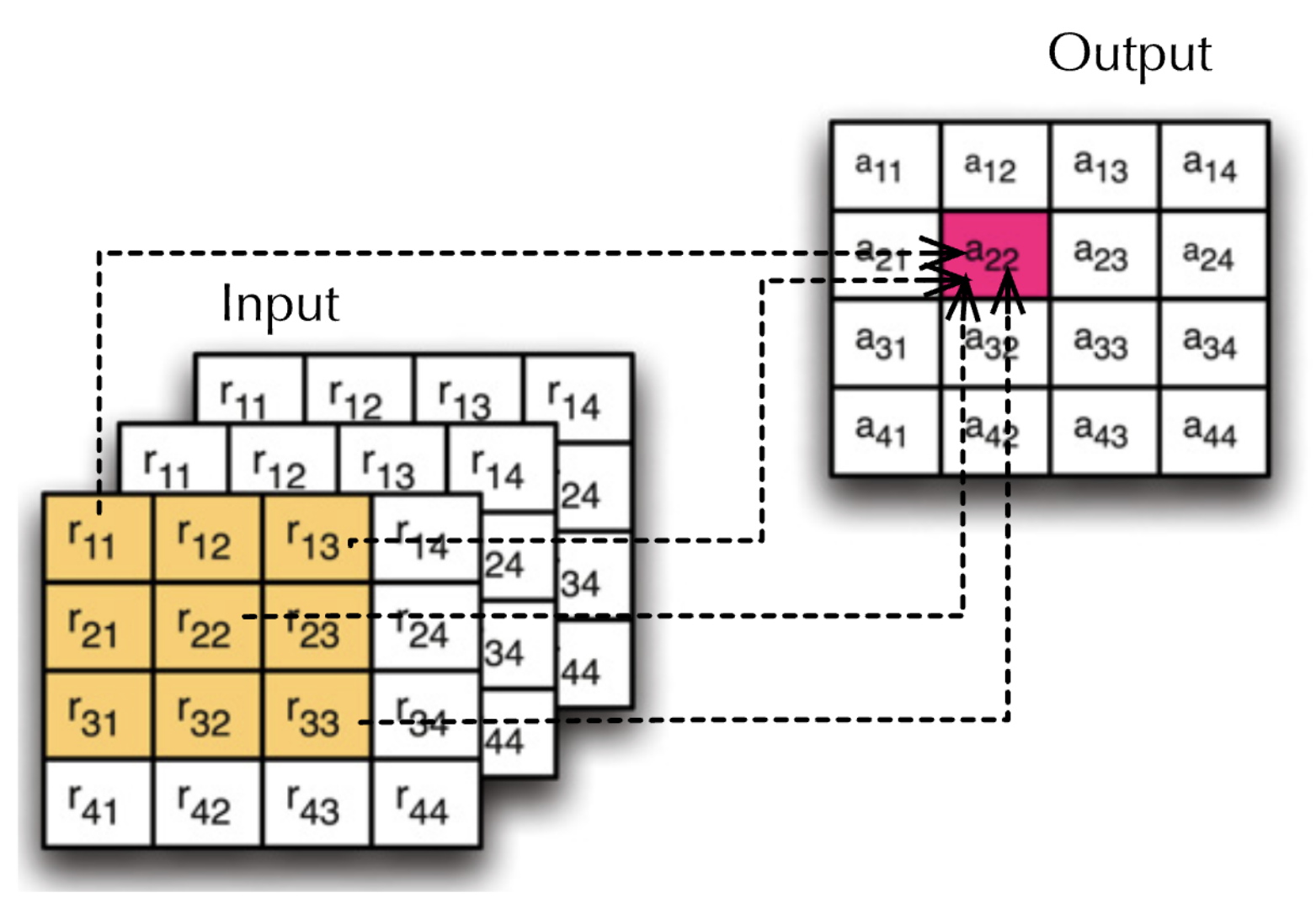} 
        \caption{Spatial neighbor-based processing}
        \label{fig:neighbor-processing}
    \end{subfigure}\hfill
    \begin{subfigure}{0.3\textwidth}
        \centering
        \includegraphics[width=\textwidth]{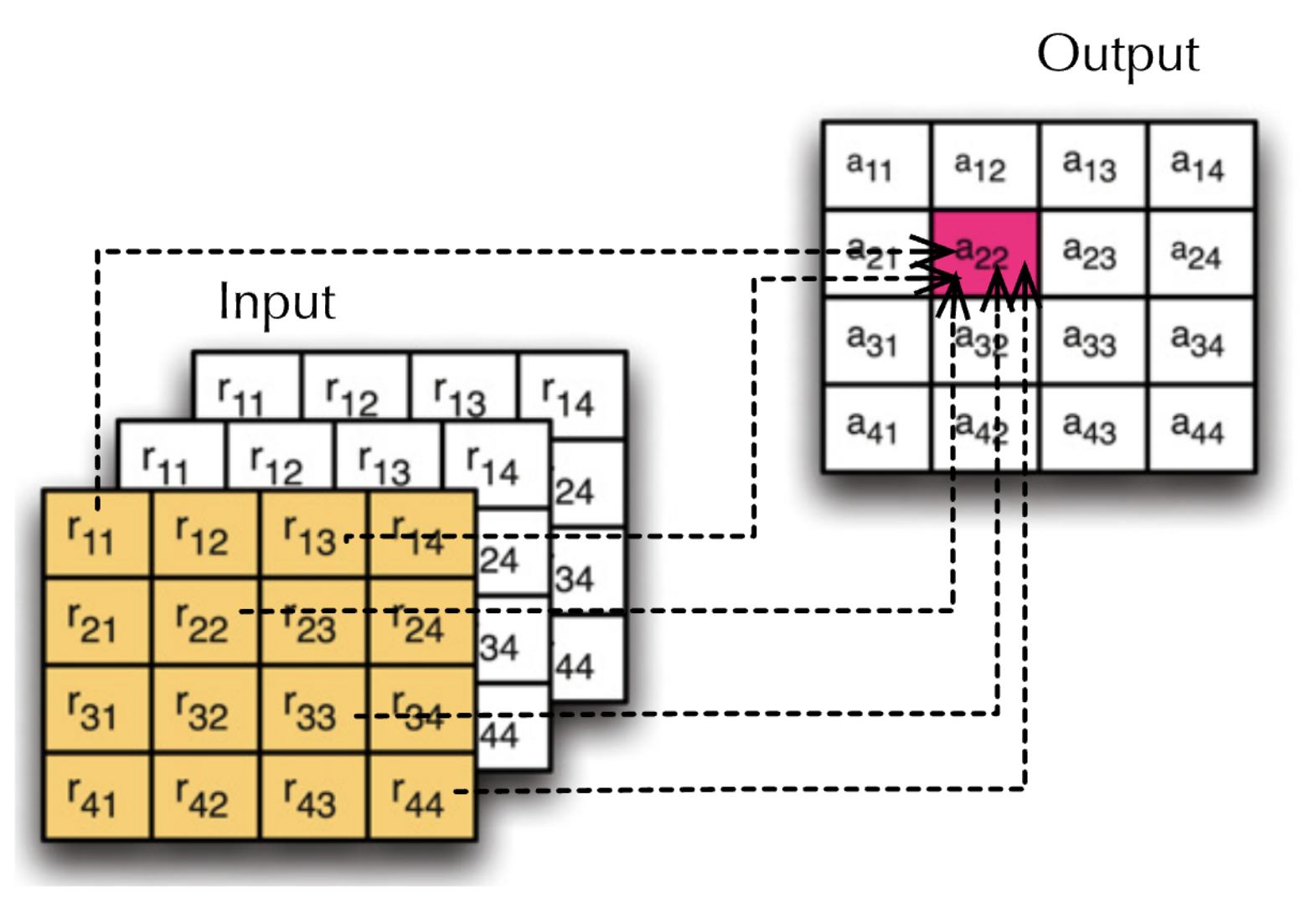} 
        \caption{Band-related processing}
        \label{fig:band-processing}
    \end{subfigure}
    \caption{Data partition on mosaicing tasks~\citep{ma2015remote}}
    \label{fig:data-access}
\end{figure}

    EOBD applications often exhibit diverse and non-contiguous data access patterns due to the pixel processing dependency with its spatial neighborhood or with data from different spectral bands as illustrated in Figure \ref{fig:data-access}. For instance, object detection workflows, which emphasizes regional dependencies, often requires numerous small data blocks dispersed within a single image file in a logical I/O operation. Conversely, band-dependent computations like vegetation health analysis involve simultaneous access to small data regions across multiple spectral band files. These patterns present a dual-edged impact: while their fragmented nature enables parallel processing across distributed worker nodes (e.g., concurrent reads of distinct blocks or bands), they introduce coordination overhead from fine-grained I/O requests and inter-task dependencies. For example, synchronizing multi-band computations or merging region-based partial results demands careful orchestration to avoid bottlenecks. Consequently, a notable challenge arises in handling the imbalance between compute and I/O intensive operations to minimize the bottleneck. Furthermore, in hybrid CPU-GPU architectures, the contention between compute and I/O operations becomes particularly critical.

    This condition can be a driver of potential energy inefficiency. For instance, a high-performance computing unit (e.g. GPU), may be forced into an underutilized state while waiting for the storage or network system to deliver required data. During this waiting time, the processor consumes static power without performing its computational task.

    \subsection{Interdependent processing workflows}
    
\begin{figure}
    \centering
    \includegraphics[width=0.5\textwidth]{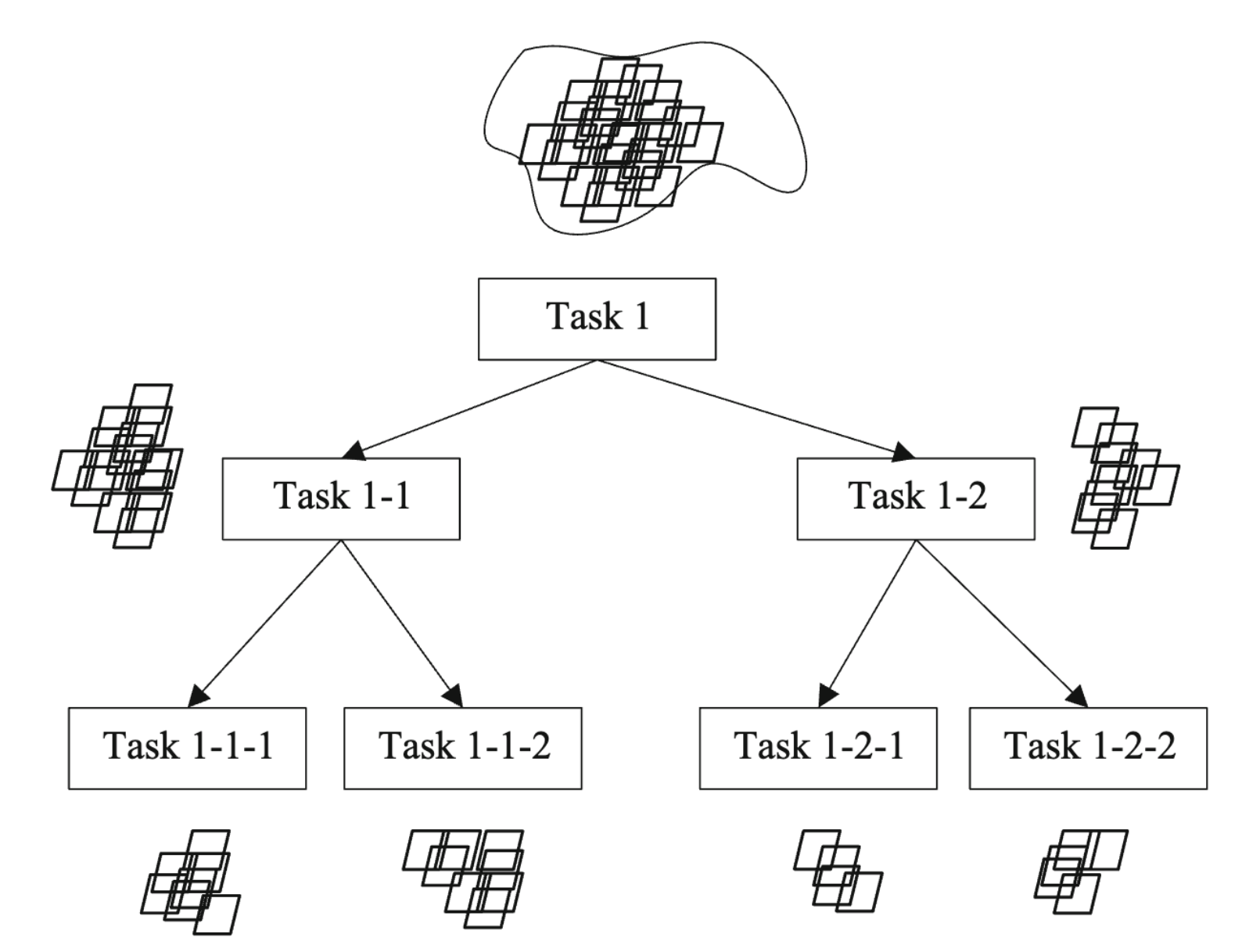}
    \caption{Data partition on mosaicing tasks~\citep{chen2015review}.}
    \label{fig:mosaic}
\end{figure}

    An EOBD processing workflow is often composed of numerous interdependent tasks, where outputs from one stage serve as inputs to subsequent stages, as illustrated in Figure~\ref{fig:mosaic}. For example, given the large area of interest (e.g., country or continental-scale), mosaicing workflows typically rely on parallel processing frameworks to stitch multiple EO tiles into single large-size composite image. Here, task interdependencies enforce strict execution order: atmospheric correction must precede tile alignment, and overlap validation must complete before seamline computation or pixel blending can begin. 
    
    Consequently, tasks are often blocked until their predecessors complete, thereby creating scheduling bottlenecks which may lead to resource under-utilization across computing cluster. While a few nodes are busy with a prerequisite task, the majority of nodes allocated to specific jobs might sit idle. If the workflow is not carefully scheduled, this idleness can represent a power drain as provisioned computing nodes consume energy without performing useful work.

\section{Landscape of cloud-based services for EOBD processing} \label{section:landscape}
    As the EOBD domain transitions toward cloud-native workflows, it is important to understand the technical demands and existing ecosystem that enable this shift. This section explores them in two main parts: 1) requirements driving the adoption of cloud-based EOBD processing, and 2) the evolving landscape of cloud services for EOBD processing, including proprietary and open-source platforms.
    \subsection{Requirements of EOBD cloud services}
    Reflecting on the unique characteristics of EOBD mentioned in Section~\ref{section:eobd_characteristics}, cloud-based EOBD processing has a number of specific unique requirements compared to other cloud-based services. First of all, the cloud platform should accommodate efficient EOBD access patterns, that allow users to access only the specific data portions needed rather than entire files~\citep{Yang2017A}. This requirement has led to the development of cloud-optimized raster formats, which organize data into hierarchical tiles enabling selective access based on spatial and temporal parameters\citep{huang2018cloud}. The Cloud Optimized GeoTIFF (COG), a prominent example, standardized the indexing mechanisms on GeoTIFF structure that facilitate direct access to specific data segments without requiring complete file download\citep{iosifescu2021cloud}. Moreover, the platform should support data storage at multiple resolutions, which is crucial for efficiently viewing and processing spatiotemporal data cubes~\citep{Zhao2019Spatiotemporal}. Support for interoperability is another important requirement due to the diversity of EO data formats (e.g. NetCDF, HDF5, GeoTIFF) and the need to process datasets from different EO missions. To address this, cloud platforms would benefit from adopting cloud-native geospatial initiatives, such as the SpatioTemporal Asset Catalog (STAC) for metadata standardization~\citep{hanson2019open}. These standards enable federated discovery and processing of EOBD datasets across public cloud providers by abstracting storage complexities through RESTful APIs. 

    Distributed processing capability is required for the parallel processing of EO data chunks, which significantly enhances the EOBD processing efficiency and reduces processing times for large datasets~\citep{Xing2019Intelligent}. Regarding the data processing model, EOBD aligns more closely with batch processing approaches, primarily due to the vast amounts of spatial and temporal data collected over extended periods~\citep{di2023big}. Batch processing, by design, is adept at handling such large datasets by organizing and processing data in comprehensive, discrete chunks, which  allows for more efficient resource utilization and the ability to conduct complex analyses that are computationally intensive. This method is particularly suited to analyze historical trends and patterns over time, a common requirement in many EO applications. While stream processing excels in real-time applications like emergency response~\citep{sun2019real}, most EOBD tasks (e.g.\ climate modeling, vegetation mapping) require a comprehensive analysis that batch processing provides. This reliance on batch processing requires auto-scaling capabilities to dynamically provision or decommission resources based on workload phases to avoid bottlenecks during peak demands and minimize the amount of deployed resources during idle periods.

    \subsection{Existing cloud-based solutions}
     Google introduced its Earth Engine (GEE) service in 2010 to provide a ready-to-use cloud-based EOBD processing service that integrates with petabytes of satellite data~\citep{gorelick2017google}. This initiative was later followed by Microsoft with the launch of the Planetary Computer (MPC) focused on open environmental datasets and providing scalable compute resources~\citep{lukacz2022data}. In 2016, Sinergise launched Sentinel Hub service to provide set of APIs and prebuilt routines to perform on-demand analyses (e.g., NDVI at continental scales) of EO data, particularly from Sentinel missions~\citep{gomes2020overview}. Separately, ESA funded the OpenEO project in 2021, which offers a standardized and open API for EO processing across multiple backends ~\citep{schramm2021openeo}. Architectural transparency varies significantly across these services. GEE, MPC, and Sentinel Hub operate as proprietary systems that are highly integrated with their underlying infrastructure. In contrast, OpenEO adopts an open-source model, decoupling its API specification from backend implementations, which are mostly open-source.

\begin{figure}
    \centering
    \includegraphics[width=0.5\textwidth]{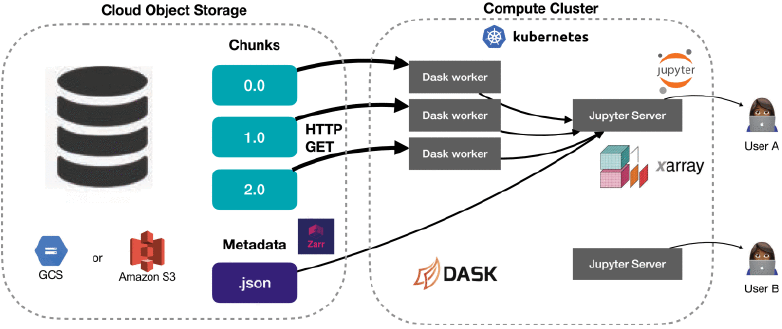}
    \caption{Pangeo System Architecture~\citep{abernathey2021cloud}}
    \label{fig:pangeo}
\end{figure}

    While OpenEO emphasizes standardization through a unified API for multiple backend compatibility, Pangeo provides a comprehensive open architecture for big data processing platforms, especially for the EO domain, as illustrated in Fig.~\ref{fig:pangeo}~\citep{abernathey2021cloud}. The platform implements Kubernetes for container orchestration, integrated with Dask Gateway for dynamic provisioning and scaling of computational resources across Kubernetes clusters~\citep{medel2018characterising}. This architecture can be deployed across major public cloud platforms or on-premise infrastructures. The system utilizes cloud object storage services (such as Amazon S3) for large-scale optimized geodata formats such as COG and Zarr. Pangeo's computational framework integrates Dask for parallel computing capabilities with xarray for multi-dimensional array manipulation~\citep{hoyer2017xarray}. The platform facilitates interactive computing through JupyterHub to enable multi-user collaborative environments.

\section{Approaches to energy efficiency in EOBD processing}\label{section:approach}
In this section, we review existing literature on energy efficiency in EOBD processing. We begin by outlining our methodology for literature collection, filtering, and analysis. Subsequently, we examine the reviewed works, which we have systematically categorized according to their relevant layers of concern within big data systems.  
\subsection{Literature review}

We conducted a literature review to get a comprehensive overview of existing energy efficiency approaches for cloud-based big data processing, with particular focus on the EO domain. Given the emerging nature of this field, our aims was to map the broad extent of research activity, identify key concepts, and uncover critical research and implementation gaps.

\subsubsection{Information sources and search strategy}
The Scopus database was selected as the primary information source due to its comprehensive coverage of peer-reviewed literature in computer science, engineering, and related fields relevant to cloud computing, big data, and EO. The search strategy focused on retrieving English-language articles published between January 2014 and December 2024. These language and publication window filters were applied directly within the Scopus search interface to define the initial set of potentially relevant records. 

Our search strategy combined keywords related to core concepts using boolean operators: (1) energy efficiency, (2) concepts related to either big data processing or cloud computing, and (3) EO. Like this we could capture relevant EO studies even if they did not strongly emphasize 'big data' or 'cloud' aspects simultaneously using keywords. Example search terms for each concept included are presented in Table \ref{tab:domain_keywords}.

\begin{table} 
  \caption{Concepts and associated terms}
  \label{tab:domain_keywords}
  \centering 
  \begin{tabularx}{\columnwidth}{ l X }
    \toprule 
    \textbf{Concept (AND operator)} & \textbf{Terms (OR operator)} \\
    \midrule 
    Energy efficiency & energy efficien*, power efficien*, green comput* \\
    \addlinespace 
    Big data on cloud computing & computing, processing, big *data, large scale, distributed \\
    \addlinespace
    Earth observation & earth observation, remote sensing, satellite, geospatial, raster \\
    \bottomrule 
  \end{tabularx}
\end{table}

\subsubsection{Eligibility criteria and selection process} 
Studies retrieved from the search were subjected to a multi-stage screening process based on predefined eligibility criteria. To be included, studies had to meet all the following conditions:
\begin{enumerate}
    \item Address energy efficiency or power consumption reduction as an explicit goal, component, or reported result of the data processing workflow.
    \item Dealt with or were applicable to big data processing within EO domain.
    \item Focus on computational processing of EO data within cloud environments after data acquisition and ingestion from original sources (satellites, sensors, etc.).
    \item Appear in a peer-reviewed journal or conference proceeding.
\end{enumerate}
On the other hand, studies were excluded if they met any of the following conditions:
\begin{enumerate}
    \item Focused solely on the energy consumption of end-user devices, IoT sensors, or data acquisition hardware (e.g., optimizing battery life for ground sensors) rather than the computational processing of the data.
    \item Addressed the energy efficiency of physical spatial objects being observed (e.g., calculating the energy footprint of buildings or cities using EO data) instead of the energy consumed during the processing of that data.
    \item Based entirely on simulation results without validation or clear applicability to real-world systems or testbeds.
\end{enumerate}

\begin{figure}
    \centering
    \includegraphics[width=0.5\textwidth]{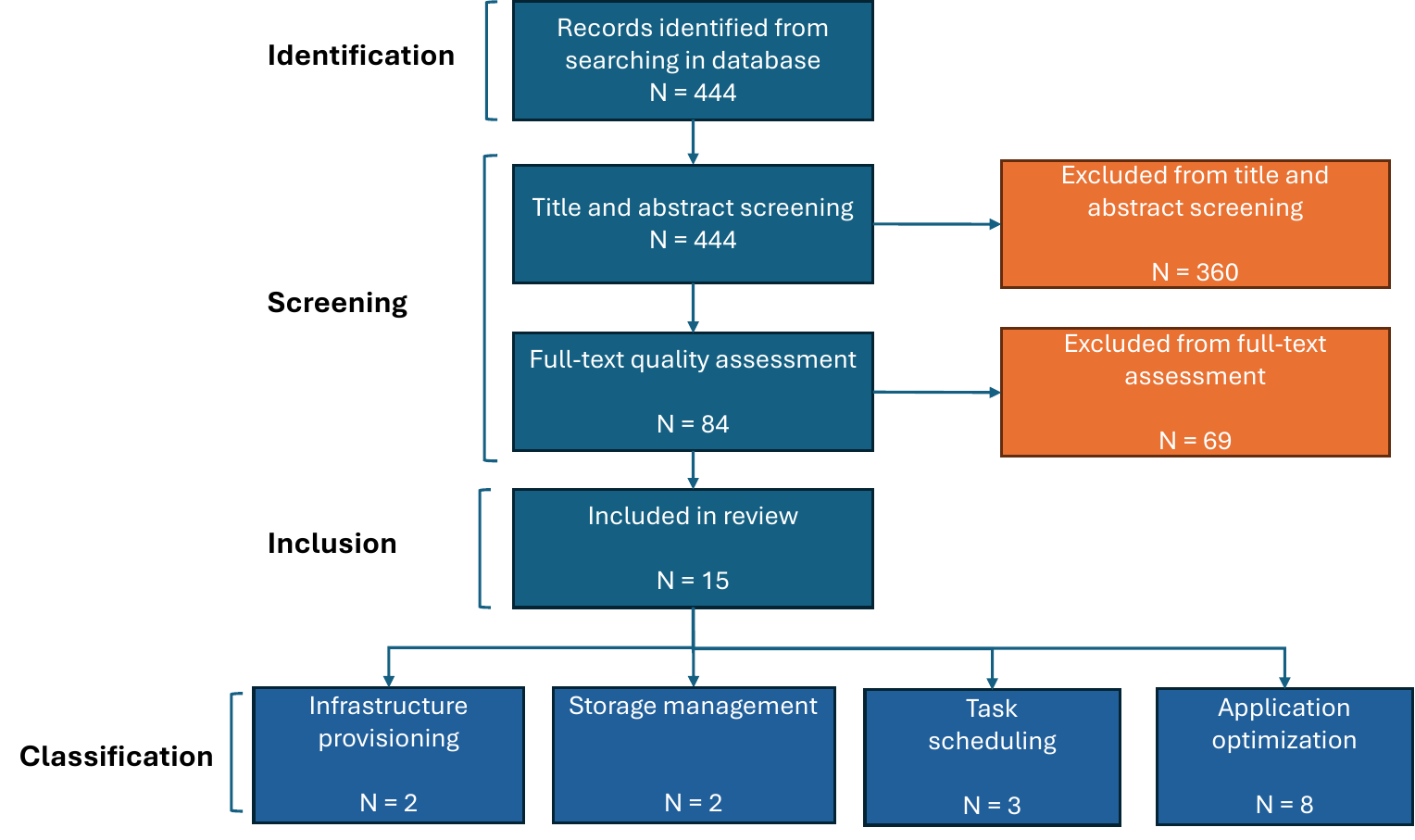}
    \caption{Flowchart of Literature Review Methodology}
    \label{fig:flowchart}
\end{figure}

At the first phase, titles and abstracts were reviewed against the eligibility criteria. Records clearly not meeting the criteria were excluded. Records deemed potentially relevant or unclear proceeded to the next stage. This step explicitly filtered based on relevance to energy efficiency, EOBD, and cloud processing context. On the second phase, the full texts of potentially relevant articles were retrieved and thoroughly assessed against all inclusion and exclusion criteria. For instance, studies initially seeming relevant might be excluded here if found to be solely simulation-based or focused on non-processing energy aspects. Reasons for exclusion at this stage were documented. At the end, studies passing the full-text review formed the final dataset for analysis. A flow diagram in Fig.~\ref{fig:flowchart} documents this process which shows the flow of records through searching, screening, eligibility assessment, and final inclusion~\footnote{The search query and complete list of reviewed articles are available at https://doi.org/10.5281/zenodo.17158339}.

\subsubsection{Synthesis}
The information extracted from the selected studies is synthesized in the next sub-sections. To provide a structured narrative of energy efficiency efforts across the data lifecycle, the identified strategies and tools were categorized based on the NIST Big Data Reference Architecture framework into four main thematic containers: infrastructure, storage management, distributed processing, and application~\citep{chang2019nist}.

    \subsection{Infrastructure provisioning}
    The infrastructure set-up of a cloud-based big data system typically involves multiple worker nodes that collaborate within a cluster to manage and process vast amounts of data. Therefore, the overall power consumption of a cluster is the accumulation of overall processing units alongside with its additional operational equipments (e.g. network, cooling, lighting) which consists of both static and dynamic power usage~\citep{ismail2020computing}. This section examines approaches spanning processor architecture selection and specialized hardware configurations.

    At the foundational level, processor architecture selection presents an aspect for energy optimization in EOBD processing.~\cite{tyutlyaeva2017energy} conducted an energy consumption analysis of night fire detection algorithms across three different Intel processor architectures: Haswell (2013), Broadwell (2014), and Knights Landing (KNL, 2016). Their study examined the correlation between the number of MPI processes and OpenMP (OMP) threads and total energy consumption on each architecture to identify the most energy-efficient MPI/OMP configuration. From their observation, Haswell delivered the best overall energy efficiency, while Broadwell processors achieved the highest energy efficiency at the CPU level, and KNL demonstrated the lowest overall energy efficiency among the three architectures. However, the Broadwell advantage was negated by high DRAM energy consumption which highlights that analyzing CPU energy consumption in isolation provides an incomplete picture of system efficiency. A particularly important finding was that I/O stages consumed significantly more energy than the core image processing stage across all architectures, with this disparity being most visible on the KNL architecture.
    
    Building upon processor architecture choices, researchers have also explored specialized hardware configurations to optimize energy consumption in EOBD processing. ~\cite{zhang2015tiny} investigated the performance and energy efficiency of big spatial data processing on a tiny GPU cluster built with Nvidia Tegra K1 (TK1) System-on-Chip (SoC) boards. They tested two real-world spatial join applications: a data-intensive task involving 170M taxi pickup points with 38K census polygons and a compute-intensive task processing species occurrence points (10M/50M) with 14K ecological polygons. Their findings revealed that ARM CPUs on TK1 were 1.48× more energy-efficient than Intel CPUs in standalone settings. However, TK1 GPUs underperformed compared to desktop/server GPUs (e.g., 24× slower than GTX Titan), resulting in lower overall efficiency.

    While these studies underscore the significant impact of hardware selection on energy efficiency, a notable gap remains in the literature regarding dynamic infrastructure provisioning strategies, particularly within cloud environments. Specifically, there is limited research exploring how to strategically select, deploy, and scale cloud resources to optimize the energy-performance trade-off for various EOBD workloads.
    
    \subsection{Storage management}

    Efficient storage management plays a crucial role in reducing energy consumption when processing EOBD. Current approaches to achieve energy efficiency in this domain can be categorized into two main strategies: hot-cold zoning and precomputed statistics.
    
    Regarding hot-cold zoning,~\cite{ye2016energy} proposed an energy-efficient strategy for optimizing storage and access of remote sensing data in Hadoop Distributed File System (HDFS). Their approach divides remote sensing images into a grid of data blocks and creates an access frequency matrix to rank these blocks. Based on this ranking, blocks are grouped according to their access patterns, with high-frequency groups prioritized for storage in nodes with the most residual space. Non-grouped blocks are stored randomly across the system, proportional to each node's available storage capacity. A key energy-saving feature of this strategy is its ability to migrate non-grouped data during periods of low computational load which allows underutilized nodes to be powered down, thereby reduce the overall energy consumption. While the strategy shows promise, its validation was limited to simulation, which highlights the need for empirical verification in a working distributed system. 
    
    In contrast to zoning approach,~\cite{li2023histogram} developed a histogram cube (HCube) model that leverages precomputed spatiotemporal aggregations of EO data. This approach represents aggregation tasks as multidimensional histogram-based lookups, with frequency histograms of EO data (such as NDVI values) stored in sparse in-memory cuboids. The model handles distributive operations (e.g., sum, count) and algebraic functions (e.g., mean) through direct computation, while approximating holistic functions (e.g., median, variance) from the stored histograms. Performance evaluations demonstrated that the proposed approach achieved up to 2× response performance and energy efficiency over existing XCube implementation. Moreover, the HCube approach exhibited lower CPU and memory utilization compared to real-time calculation methods. However, this efficiency comes with a trade-off in accuracy due to histogram approximation and grid distortions.

    \subsection{Resource allocation and task scheduling in distributed processing}
    The execution plan of a distributed application is often represented as a Directed Acyclic Graph (DAG). In this representation, each vertex corresponds to a specific operation within the application and edges depict the direction, order, and dependencies of the data and execution flow~\citep{salloum2016big}. Based on this model, distributed processing frameworks take on parallelizing these operations into several tasks, which are then distributed across worker nodes by a scheduling module~\citep{hasan2021survey}.

    \cite{sun2020multiobjective} presented an energy-aware task scheduling method for hyperspectral image classification using the Multi-Objective Immune Algorithm (MOIA) that simultaneously minimizes execution time and energy consumption. Their approach develops a parallel implementation of fusion-based hyperspectral classification on Apache Spark, where MOIA mimics immune system mechanisms such as selection, mutation, and recombination to balance trade-offs between competing objectives. The evaluation was performed on a computing cluster with up to 32 worker nodes. By iteratively searching for optimal solutions through candidate evaluation, their method achieved 36.21\% improvement in energy efficiency compared to standard Spark parallel processing algorithms.
    
    On the resource allocation approach,~\cite{senapati2024ers} developed an energy-efficient heuristic resource allocation approach for DAG-based tasks. Their method maximizes energy savings by selecting task-to-processor frequency assignments based on computed task priorities using rank-based metrics that favor tasks with higher computational and communication overheads. The approach utilizes Effective Start Time (EST) and Effective Finish Time (EFT) metrics to allocate tasks while incrementally adjusting task frequencies to meet deadlines with minimal energy consumption. Experimental results through simulation showed that their approach outperformed existing methods such as global and simple static power management techniques with 28-84\% higher energy savings. However, a notable trade-off exists as run-time increases with task count which potentially limits its practical scalability for very large DAG-based workflows.
    
    \subsection{Application-specific computational optimization}
    Application-specific computational optimization represents a targeted approach to energy efficiency in EOBD processing by implementing specialized algorithms on hardware accelerators such as GPUs and FPGAs. These optimizations exploit the inherent parallelism capabilities of such hardware to achieve energy efficiency improvements for specific computational tasks.

    \cite{baumeister2022fast} demonstrated this approach by implementing the emissivity growth approximation (EGA) method on GPU architectures to accelerate radiative transfer calculations. They restructured the JURASSIC radiative transfer model using CUDA, focusing on ray-tracing optimizations, memory access patterns (e.g., coalesced memory access), and kernel fusion techniques to minimize data transfers between CPU and GPU. When tested on NVIDIA Tesla V100 and A100 GPUs for both limb and nadir observation geometries, their GPU implementation achieved promising results: 9× higher energy efficiency and 14x faster runtime compared to an optimized CPU version running on Intel Xeon Gold 6148 processors.

    In the domain of hyperspectral image processing,~\cite{ortiz2018runtime} developed a data-parallel version of the Fast UNmixing (FUN) algorithm for hyperspectral linear unmixing. Their approach implemented a hardware-accelerated solution using the ARTICo framework on reconfigurable FPGA-based systems. The FUN algorithm was adapted for data parallelism by splitting hyperspectral images into blocks, where partial end-members from each block are iteratively reduced to produce final results. Their experimental setup employed a cluster of Zynq-7000 SoPC nodes to emulate distributed satellite processing, with Message Passing Interface (MPI) managing data distribution across nodes. Experimental results showed that using four accelerators improved performance by up to 8.8× over software-only implementation while reducing energy consumption by almost 50\%.

    For remote sensing scene classification,~\cite{zhang2020efficient} implemented a quantized version of the Improved Oriented Response Network (IORNN) on FPGA accelerators. Their approach quantized the model to 8-bit fixed-point representation using quantization-aware training techniques. The implementation optimized convolutional and fully connected layers through parallelism (utilizing 64 processing elements), data reuse strategies, and operation fusion (e.g., merging quantization and ReLU activation functions). They minimized off-chip memory access through strategic buffer placement and employed custom data reordering to match FPGA bit-width constraints. To evaluate the power efficiency of their implementation, the authors used Giga operations per second per watt (GOPS/s/W) as a metric, where higher values indicate better energy efficiency. The proposed approach achieved an energy efficiency of 33.16 GOP/s/W which outperforms both CPUs by around 174× (0.19 GOP/s/W) and GPUs by around 11× (2.88 GOP/s/W).

    Similarly,~\cite{rajesh2024optimizing} proposed an FPGA-based architecture specifically optimized for hyperspectral image enhancement tasks. Their design utilized on-chip Block RAM for frequently accessed data while employing off-chip DDR3 memory for storing the complete dataset. This hierarchical approach minimized memory access latency and reduced overall power consumption. Experimental results demonstrated that the FPGA implementation consumed only 0.5 Joules per processed image, substantially outperforming both GPU (13 J) and CPU (22.5 J) implementations. These findings highlight the potential of FPGA-based solutions to achieve significant energy savings in computationally intensive EO data processing tasks, particularly for applications requiring high-dimensional data manipulation and parallel processing capabilities.

    In another study,~\cite{zhang2023accelerating} demonstrated the utilization of high bandwidth memory (HBM) on modern FPGA devices to accelerate the execution of Graph Neural Network (GNN) computational kernels, including feature aggregation and feature update operations. The HBM provided the necessary capacity (8 GB) and bandwidth (460 GB/s) to feed the processing elements, significantly outperforming the limited on-chip memory alternatives. Their evaluation revealed that the FPGA implementation was vastly more energy-efficient, consuming only 0.05 J/image, which represents 36.2× better efficiency than CPU implementation (1.81 J/image) and 7.35× better efficiency than GPU implementation (0.246 J/image).

    The field has also explored approximate computing paradigms to achieve energy efficiency through intentional precision reduction.~\cite{jia2023energy} proposed an image change detection method based on frequency analysis using an approximated Discrete Cosine Transform (DCT) implemented on customized processors. Their approach leverages the observation that many image processing applications, including change detection, exhibit inherent tolerance to minor computational errors. This tolerance enables the application of approximate computing strategies where computational precision is intentionally reduced to decrease hardware complexity and power consumption. The proposed approximated DCT demonstrated substantial hardware resource reductions compared to exact DCT implementations: 60.86\% reduction in circuit area and 64.77\% reduction in power consumption while maintaining acceptable accuracy for change detection tasks.

    Exploring neuromorphic computing paradigms,~\cite{kadway2023low} investigated Spiking Neural Networks (SNNs) to enable energy-efficient, real-time cloud cover detection on the on-board devices as proof-of-concept for in-satellite data processing. Their approach converted a pre-trained Convolutional Neural Network to an SNN using BrainChip's Akida platform. The implementation featured a two-stage hierarchical cloud cover detection model: a coarse stage operating on 64×64 pixel patches followed by a fine-grained classification on 8×8 pixel patches. The first stage achieved 35× lower energy consumption and 3.4× faster latency compared to a Jetson TX2 platform, while the second stage demonstrated more efficient results with 230× lower energy consumption and 7× faster latency.

    Building upon the neuromorphic computing approach, SNNs have emerged as promising alternatives to traditional Artificial Neural Networks (ANNs) due to their event-driven computation model. SNNs consume energy only when neurons fire, which leads to sparse activity patterns and substantially reduced power requirements, particularly when deployed on specialized neuromorphic hardware. However, this spike-based communication mechanism can introduce higher processing latency compared to the parallel, instantaneous computations of ANNs, especially for tasks requiring immediate, continuous output. To address this limitation,~\cite{li2024spatio} proposed a spatiotemporal pruning method that integrates spatial feature compensation with temporal latency reduction to enhance ultra-low-latency SNN performance for EO scene classification. Their approach employs a knowledge distillation strategy where a deeper, more capable network is temporarily trained to transfer learned knowledge to a smaller, final network. Experimental results demonstrated that SNN energy consumption was 10× to 100× lower than equivalent ANN implementations, with the reduction factor dependent on network depth and complexity.

    Despite these promising results, application-specific computational optimizations face a common limitation: they typically focus on optimizing a single specific task for a particular type of hardware accelerator. This narrow focus limits their generalizability across different EO applications and hardware platforms. In this case, re-engineering efforts are required to adapt new tasks or deploy on different hardware configurations.
    
\section{Gaps of energy efficiency in EOBD processing}\label{section:challenges}

Based on the characteristics of EOBD processing presented in Section~\ref{section:eobd_characteristics}, the landscape of EOBD services in Section~\ref{section:landscape}, and various energy efficiency approaches in general big data in Section~\ref{section:approach}, we identify several challenges to realize energy efficient EOBD processing.
    
    \subsection{Redundant high volume data}
    
    The EOBD life cycle typically begins with satellites capturing raw imagery, followed by transmission to ground stations and subsequently to data archival facilities where the data undergoes initial processing including radiometric and geometric corrections. The processed data is then distributed to various data provider storage services for broader access, additional pre-processing, and analysis~\citep{Xu2022Cloud-based}. This practice creates redundancy, as nearly identical pre-processing operations are often repeated independently by competing data providers to meet the needs of analysis ready data. Once processed, data replication escalates energy demands further: providers maintain multiple copies of multi-terabyte products (e.g., Sentinel-2 Level-1C) across geographically distributed tiers~\citep{nguyen2010differentiated, PlanetarySTAC2024}, such as edge nodes for low-latency environmental monitoring (e.g. wildfire) or centralized archives for decades-old climate records like the Landsat. While this redundancy aligns with user needs to ensure rapid access for time-sensitive disaster response or regulatory compliance with open data policies, it multiplies energy costs from storage, including its cooling and server operation, and cross-regional transfers.

    Much of the existing research on energy-efficient data storage for EOBD has focused on intra-system optimizations, such as hot-cold data zoning~\citep{ye2016energy} or efficient data access with precomputed statistics~\citep{li2023histogram}. While these techniques offer promising gains within a single data repository, their overall impact is constrained by the broader, inter-system challenge of data duplication. This limitation suggest that the future work should extend beyond refining storage-level techniques to explore new system-level architectural models, such as data federation or collaborative stewardship models, that can reduce large-scale redundancy across the EOBD ecosystem. 

    \subsection{Limited information on actual energy consumption from cloud provider}
    
    Cloud providers typically offer monitoring tools that enable users to track various performance and resource usage metrics, including CPU utilization, memory usage, and I/O activity. AWS, for example, provides CloudWatch, a comprehensive monitoring service with a two-tier pricing model: basic (free) and premium (paid)~\citep{AWSCloudWatchPricing}. The free tier includes fundamental metric measurements at 5-minute intervals for core services such as EC2 (computing resources), EBS (storage), S3 (object storage), and RDS (database services)~\citep{diagboya2021infrastructure}. The premium tier offers enhanced capabilities, including more granular 1-minute interval measurements and detailed custom metrics. Similar monitoring services are available from other major cloud providers, such as Microsoft Azure Monitor and Google Cloud Monitoring. Furthermore, organizations can augment these native monitoring capabilities with popular open-source solutions like Prometheus for metrics collection and Grafana for visualization, creating more comprehensive monitoring ecosystems~\citep{sukhija2019towards}.

    Despite these advanced monitoring capabilities, cloud providers still provide limited transparency regarding energy consumption metrics for workloads running on their infrastructure~\citep{moghaddam2021metrics}. While AWS has introduced the Customer Carbon Footprint Tool to help organizations track their carbon emissions, this calculator has several limitations. Although it provides monthly carbon footprint estimates broken down by AWS services (e.g., EC2, S3, RDS), it lacks granularity at the instance and task level~\citep{arora2023towards}. This means that organizations cannot measure the energy consumption of specific workloads, individual computing instances, or particular computational tasks. The tool also relies on average energy consumption and carbon intensity factors rather than real-time measurements. This transparency limitation extends to specialized EO cloud platforms, as prominent services like GEE and Pangeo Cloud also lack comprehensive energy consumption measurements for their operations.

    The implications of this transparency gap become evident when considering the validation of existing research. The energy efficiency gains reported in studies using specialized hardware, such as works on GPU by~\cite{baumeister2022fast} and FPGA by~\cite{zhang2023accelerating}, are derived from controlled, on-premise experiments where power can be directly measured. In a public cloud environment, a researcher or organization has no practical way to verify if deploying a similar algorithm on a cloud-based GPU or FPGA instance actually yields the expected energy savings.
    
    \subsection{Lack of reliable and reproducible benchmarking on energy efficiency}
    
    The development of standardized frameworks and tools for benchmarking energy efficiency in cloud-based EOBD processing remains understudied~\citep{bhawiyuga2025energy}. While numerous benchmark frameworks exist in the EO domain, they predominantly prioritize functional performance metrics such as response time, scalability, or algorithmic accuracy. For example, a framework was proposed to benchmark server-side EO data processing service based on Rasdaman DBMS with the focus on evaluating the response time~\citep{karmas2015benchmarking}. In other work, Gomes et al presented a qualitative comparison of several cloud-based EO big data services (e.g. OpenEO, GEE, pipsCloud, etc.) in term of their data, processing, and infrastructure abstractions as well as their processing and storage scalability~\citep{gomes2020overview} while neglecting the energy efficiency as important metrics. Similarly, existing datasets such as Bigearthnet~\citep{sumbul2019bigearthnet}, DIOR~\citep{li2020object}, and RSI-CB~\citep{li2020rsi} are designed to benchmark classification or segmentation accuracy.

    This absence of a reproducible energy-aware benchmarking tools and standardized datasets creates several limitations. First, it makes it impossible to perform a fair, apples-to-apples comparison of the energy efficiency of different solutions. For example, the 50\% energy reduction achieved by \citet{ortiz2018runtime} for hyperspectral unmixing on an FPGA more cannot be fairly compared with the 10-100x reduction reported by \citet{li2024spatio} for scene classification using SNNs. Without a common set of reproducible testing workflows, datasets, and energy measurement protocols, these results remain isolated data points. Furthermore, current benchmarking lack granular energy consumption profiles for EO workflows, particularly for atomic tasks such as data ingestion (I/O-intensive), compute-heavy operations (CPU/GPU-intensive), or network-bound processes. For instance, there are no standardized datasets that isolate energy consumption during raster algebra operations versus spectral index calculations. Without such scenario, researchers cannot identify the impact of different types of workflow to the energy consumption for further optimization. Most critically, energy consumption is rarely included as a measurable parameter in EOBD workflow design.

    \subsection{Lack of energy-awareness in cloud infrastructure orchestration strategy}
    
    Cloud computing has evolved from full virtualization to containerization, particularly in big data processing where scalable yet flexible infrastructure is crucial. Container orchestration plays a vital role in resource allocation management, automatic scaling, and simplified deployment of complex data processing pipelines across machine clusters. Kubernetes has become the de facto standard for container orchestration and sees wide adoption in various cloud-based big data processing scenarios, including EOBD platforms like Pangeo~\citep{munteanu2024data}.

    In Kubernetes' architecture, processes and their dependencies exist within pods, which serve as the fundamental building blocks of applications~\citep{luksa2017kubernetes}. The Kubernetes scheduler assigns these pods to available worker nodes through a two-stage process: filtering and scoring. The filtering stage identifies suitable worker nodes based on predefined constraints, while the scoring stage evaluates these nodes with specific metrics to determine optimal pod assignment~\citep{rejiba2022custom}. The default Kubernetes scheduling policy, \textit{LeastRequestPriority}, favors nodes with the highest percentage of available CPU and memory resources to optimize resource utilization. An alternative criterion, \textit{BalancedResourceAllocation}, strives for fair resource distribution among worker nodes to prevent overburdening. To handle fluctuating workload demands, Kubernetes implements several scaling strategies: the Horizontal Pod Autoscaler (HPA) adjusts the number of pod replicas based on CPU utilization, the Vertical Pod Autoscaler (VPA) optimizes resource allocation for individual pods, and the Cluster Autoscaler modifies the cluster size by adding or removing worker nodes based on current and predicted resource demands~\citep{nguyen2020horizontal}.
    
    The work by \citet{tyutlyaeva2017energy} demonstrated that different processor architectures exhibit vastly different energy efficiencies for the similar EO processing task. A standard Kubernetes pod placement and scaling mechanisms, while effectively manage performance and resource utilization, overlook energy efficiency as an important optimization parameter~\citep{carrion2022kubernetes}. This omission proves particularly counterproductive in EOBD processing workflows, where inefficient resource allocation can result in prolonged idle node operation, redundant server activation, and suboptimal energy-per-task ratios. For example, during satellite data ingestion and analysis, a Kubernetes scheduler might distribute compute-heavy pods across underutilized nodes to maximize CPU availability, which in turn maintains multiple servers at low utilization. A further limitation exists in Kubernetes' failure to account for heterogeneous worker node configurations. Modern cloud infrastructures typically comprise mixed clusters with both high-performance but power-hungry nodes (e.g., x86 servers with multi-core CPUs) and energy-efficient alternatives (e.g., ARM-based nodes with lower clock speeds but better power-to-performance ratios). The default Kubernetes scheduling policies treat all nodes as homogeneous to prioritize resource utilization metrics like CPU/memory availability over hardware-specific energy profiles. For instance, a pod that requires sustained computational power might receive assignment to an ARM-based node, which prolongs task completion time and indirectly increases energy use due to extended runtime. Conversely, lightweight but frequent batch jobs could route to high-performance nodes, thus wasting energy when lower energy alternatives exist. This one-size-fits-all approach neglects the potential to match workloads to nodes based on their energy efficiency characteristics, such as favoring ARM-based nodes for parallelizable EO tasks or reserving high-performance nodes for latency-sensitive workloads.

    \subsection{Inefficient task scheduling in distributed processing}
    
    Distributed processing frameworks like Dask, Apache Spark, and Apache Flink manage large-scale data by partitioning it into manageable chunks and processing them in parallel across a cluster of worker nodes. ~\citep{dugre2023performance}. The task scheduler plays crucial role to efficiently distribute task execution across the cluster~\citep{daskSchedulingx2014}. Current scheduling strategies primarily focus on a fair balance among worker nodes and optimal data locality. For instance in Dask, tasks without specific requirements are initially assigned to the least busy worker based on queue length and processing capacity. For subsequent tasks that require access to previously computed data, the scheduler prioritizes data locality to enhance performance and reduce data transfer overhead. If no available workers have the required data, the scheduler considers the additional time needed for data transfer to potential candidates, and incorporates network topology and bandwidth constraints into the decision-making process.

    While current scheduling strategies effectively balance workload and data locality, they overlook energy efficiency. These strategies do not account for the energy consumption characteristics of worker nodes, which can lead to higher overall energy consumption for the entire cluster. For instance, assigning a task to a high-energy-consuming worker solely based on data locality may result in greater total energy expenditure compared to using a more efficient worker, even when accounting for data transfer costs. This oversight highlights the need for energy-aware task assignment strategies in distributed processing frameworks that optimize energy efficiency while considering the heterogeneous nature of worker nodes.

    While researchers have proposed energy-aware multi-objective scheduling algorithms \citep{sun2020multiobjective, jiang_orbit_2023}, these advanced capabilities are not standard features in distributed processing frameworks like Dask, Spark, or OpenMPI. A further challenge lies in the practicality of these advanced algorithms. Many sophisticated methods, such as those based on evolutionary computation, carry a significant computational overhead that may introduce unacceptable latency into the scheduling process itself which makes them unsuitable for dynamic, real-time decision-making. Therefore, a gap also exists in assessing the trade-off between algorithmic optimality and computational feasibility.

\section{Future research directions}\label{section:improvement}
The analysis in Section \ref{section:challenges} has identified several gaps that currently hinder energy-efficient EOBD processing in the cloud. These gaps span from the lack of energy monitoring and standardized benchmarks, to systemic inefficiencies in how the infrastructure is orchestrated and how computational tasks are scheduled. To overcome these limitations, a comprehensive approach is required that addresses multiple layers of the computational stack. Accordingly, this section outlines three interconnected research directions aimed at closing the identified gaps. First, we investigate the design and requirements of a toolkit for energy benchmarking and monitoring of EOBD processing. Second we explore key research questions related to energy-aware infrastructure orchestration. Finally, we examine the potential for developing task scheduling algorithms within distributed processing frameworks that can co-optimize for energy efficiency and computing performance.
\subsection{EOBD-specific energy benchmarking and monitoring toolkit}
    As highlighted in Section~\ref{section:challenges}, existing cloud-based EOBD platforms lack granular energy monitoring mechanisms tailored to distributed EOBD workflows. This gap is compounded by the absence of standardized benchmarks to evaluate energy efficiency in EOBD processing. To address these gaps, an integrated monitoring toolkit is needed to attribute energy consumption to individual workflows in distributed EO processing clusters. The toolkit should aggregate data from three different layers: 
    \begin{enumerate}
        \item Hardware-level sensors: CPU/GPU power via Intel RAPL or SMI, system-wide consumption via IPMI or external power meters.
        \item OS-level metrics: system and per-process CPU/memory usage, disk I/O, and network traffic.
        \item Application profiling: integration with EOBD frameworks (e.g. Dask’s diagnostic dashboards) to map the resource utilization of specific jobs.
    \end{enumerate}
    To attribute power consumption to specific workflows, the toolkit should incorporate a power model that correlates observed resource usage (e.g., CPU and GPU utilization, disk I/O, and network bandwidth) with energy consumption. This model could leverage historical sensor data and machine learning techniques to estimate per-workflow energy costs, even when multiple tasks share hardware resources. For example, during concurrent SAR processing and optical image classification, the model would disaggregate each workflow’s contribution to total cluster power by analyzing patterns in CPU/GPU utilization and task duration.

    In addition, an EO-specific energy benchmarking framework is needed which consists of EO-specific atomic tasks that mirror common processing steps in EO workflows, for example:
    \begin{enumerate}
        \item CPU-intensive operations: These include raster data resampling or atmospheric correction for Sentinel-2 imagery.
        \item GPU-intensive processes: Examples include deep learning based land cover classification inference.
        \item Disk I/O intensive tasks: Large-scale mosaicking operations test I/O throughput and parallel file system efficiency by stitching thousands of tiles into continent-scale mosaics.
        \item Network intensive procedures: Data tiling/chunking partitions massive datasets (such as country-level multi-spectral satellite images) acquired from cloud data providers into smaller chunks for distributed processing.
    \end{enumerate}
    To reflect real-world complexity, the framework should also include composite workflows that chain atomic tasks into end-to-end EOBD use cases. For instance, the phenology-based agriculture monitoring combines ingestion of multi-temporal EO images (I/O and network-intensive), NDVI computation (CPU-heavy), and time-series analysis to track start of season (memory and CPU-intensive)~\citep{zurita2019exploring}.

    \subsection{Energy-efficient infrastructure orchestration}
    
    EOBD workflows are inherently heterogeneous, with varying computational requirements and user preferences regarding deadlines. These preferences can be quantified through a "green score" ranging from urgency-driven (0.0) to energy-optimized (1.0). Users with scores closer to 1.0 prioritize energy efficiency and accept longer processing times, while those with scores approaching 0 require immediate execution regardless of energy costs. These workflows typically execute on distributed processing clusters with diverse hardware capabilities. By profiling the energy consumption of different hardware configurations during various EO workloads, detailed energy profiles can be established to inform orchestration decisions that balance performance requirements with efficiency goals.

    To operationalize energy-efficient orchestration, existing schedulers like Kubernetes can be extended to incorporate energy awareness into their decision-making processes~\citep{KubeScheduler2024}. The extended scheduler first evaluates the green score of incoming tasks to route them to worker nodes that align with user priorities. This approach enables the system to assign tasks to nodes with optimal energy-performance ratios—for example, directing lightweight preprocessing tasks to ARM processors while routing computationally intensive deep learning tasks to GPU-accelerated nodes. Consider two contrasting workflow examples running on heterogeneous clusters: For urgent workflows such as wildfire detection (low green score), the scheduler prioritizes high-performance GPU-enabled nodes to meet critical deadlines despite higher energy costs. Conversely, for non-urgent workflows like monthly vegetation index batch processing (high green score), the system can delay execution to nighttime hours when energy costs are lower and ambient temperatures cooler, or prioritize energy-efficient ARM-based nodes with lower power consumption profiles.

    A predictive autoscaling mechanism with deadline awareness can further enhance this approach by dynamically adjusting the cluster's composition. In Kubernetes, the Horizontal Pod Autoscaler (HPA) automatically scales the number of computing units based on observed metrics like CPU utilization or memory usage. While the default HPA monitors application load and adjusts resources reactively, an enhanced version can leverage historical EOBD workload data to predict resource needs, scale preemptively, and avoid last-minute over-provisioning that often leads to energy waste. For urgent workloads, this predictive HPA slightly over-provisions resources to ensure deadline compliance while still maintaining better energy efficiency than purely reactive approaches.

    Energy efficiency can be further improved through workload consolidation strategies. The implementation of bin-packing algorithms optimized for energy efficiency allows the system to group tasks with complementary resource requirements (such as CPU-bound and memory-bound processes) on the same node, which would maximize utilization while it minimizes the total number of active nodes. This approach enables the Kubernetes to power down underutilized nodes or transition them to low-power states. For instance, the system can consolidate nightly batch jobs for satellite data ingestion onto a subset of nodes, which allows others to enter sleep mode and reduce the cluster's overall energy footprint.
    
    \subsection{Multi-objective task scheduling}

    Complementary to infrastructure orchestration, further energy efficiency gains require attention to task scheduling mechanism within distributed processing systems like Dask or Spark. Current task schedulers within these frameworks typically prioritize execution speed and neglect energy consumption as the important criteria. Future research should therefore focus on the design and implementation of multi-objective strategies for task assignment specifically for EOBD workloads. These strategies must simultaneously seek the energy usage optimization, execution time (makespan), and potentially resource cost. 
    
    To enable such task assignment, accurate task-level energy models are essential. These models should estimate the energy footprint of individual EOBD computational tasks based on each task's characteristics and the specific type of worker node (CPU and GPU which are provisioned and profiled by the orchestrator) available for its execution. The development of these models could involve the creation of detailed profiles from EOBD libraries or the use of machine learning techniques that predict energy use from task metadata and system metrics.

    Equipped with task-level energy models, the next challenge becomes the selection and implementation of appropriate algorithms for task assignment. Existing approaches adapted from multi-objective optimization, such as evolutionary algorithms or reinforcement learning, have the ability to explore the complex trade-offs between energy, time, and cost, which potentially lead to near-Pareto-optimal solution. 
    However, a significant practical limitation arises from the computational complexity inherent in many of these techniques. The overhead associated with population management in evolutionary algorithms or complex policy evaluations in reinforcement learning might introduce additional latency, which are less suitable for dynamic, real-time task scheduling decisions required in operational EOBD systems. Therefore, future research must also investigate computationally lighter alternatives. This includes the development and evaluation of domain-specific heuristics (e.g. energy-aware adaptations of classic scheduling algorithms like HEFT or Min-Min) or well-designed greedy strategies. While potentially suboptimal compared to exhaustive methods, these heuristics aim to provide good, practical solutions with significantly lower computational overhead. Regardless of the chosen algorithmic approach, effective integration with the underlying orchestrator remains essential to ensure that framework-level task assignments align with the broader resource allocation policies and constraints managed at the infrastructure level.

\section{Conclusions}\label{section:conclusion}
The increasing use of cloud computing for EOBD processing has enabled advancements in large-scale environmental analysis. As the field matures, the energy consumption associated with these computational workflows should become an area of growing importance. This study examined the current state of energy efficiency in cloud-based EOBD processing to identify existing gaps and suggest potential directions for future research.
Our literature review indicates that efforts to improve energy efficiency have predominantly focused on application-specific optimizations, such as the use of hardware accelerators for particular algorithms. While effective for their intended tasks, these approaches can be difficult to generalize across the diverse range of EOBD workflows. We observe that broader, system-level aspects, including infrastructure orchestration in cloud environments and the management of large, often redundant data archives, have received comparatively less research attention. This situation is further complicated by the limited availability of energy consumption metrics from cloud providers and the lack of standardized benchmarks.
Based on these observations, several research directions appear promising. A foundational area for investigation is the development of standardized benchmarking frameworks and energy monitoring tools tailored for EOBD use cases. Such tools would facilitate empirical analysis of energy consumption as a baseline for further work. Building on this, future studies could explore improvement to infrastructure orchestration and resource allocation strategies to incorporate energy-awareness. A third possible approach involves the design and evaluation of multi-objective task scheduling algorithms within distributed processing frameworks that prioritize energy efficiency while maintaining reasonable computing performance. By outlining these critical areas, this study provides a conceptual foundation for future practical work. The proposed research directions are intended to guide the empirical studies and systems development necessary to build more energy-efficient EOBD processing.

\section{Declaration of generative AI and AI-assisted technologies in the writing process}

During the preparation of this work the authors made a limited use of Google Gemini and Claude Sonnet to improve readability. The authors take full responsibility for the content of this article.

\newpage

\textbf{Code availability section}

As this is a review article, no new software or custom code was developed for the preparation of this manuscript.

\textbf{Data availability section}

The dataset generated and analyzed for this systematic review, including the full list of screened, excluded, and included studies, is openly available in repository accessible at  https://doi.org/10.5281/zenodo.17158339

\bibliographystyle{cas-model2-names}

\bibliography{library}

\end{document}